\documentclass[twocolumn,showpacs,prl]{revtex4}
\usepackage{amsmath}
\usepackage{mathrsfs}
\usepackage{graphicx}

\begin{document}
\draft

\title{%
       Nonlocality of Dephasing in a Charge Qubit Interacting with a Quantum  Point Contact Beam Collider }%

\author{Youngnae Lee}
\author{Gyong Luck Khym}
\author{Kicheon Kang}\email{kicheon.kang@gmail.com}
\affiliation{Department of Physics, Chonnam National University,
Gwangju 500-757, Korea}

\date{\today}

\begin{abstract}

We investigate the charge-detection-induced dephasing of a charge
qubit interacting with an electronic beam collider composed of a
quantum point contact.  We report that, while the qubit is dephased by
the partitioned beam of uncorrelated electrons,
the interference of the qubit
is fully restored when the two inputs are
identically biased so that all the electrons suffer two-electron collision.
This phenomenon is related to Fermi
statistics and illustrates the peculiar {\em
nonlocality of dephasing}. We also describe detection properties for the
injection of entangled electron pairs.

\end{abstract}



\maketitle

In a two-path interferometer with a \emph{which-path} (WP) detector,
the observation of interference and acquisition of the WP
information are mutually exclusive~\cite{feynman65, wootters79,
stern90}. It has been shown that dephasing (i.e., reduction of the
interference) can be understood either as the acquisition of the
WP information or as
the back-action caused by the detector~\cite{stern90}. However, it
has been argued that the back-action dephasing is not simply
occurred as a result of the classical momentum kick in some
cases~\cite{scully91,durr98}.
Owing to the recent advances in nanotechnology, mesoscopic devices now provide
opportunities for investigating this issue.
Indeed, WP detection in quantum interferometers has been achieved
by using mesoscopic conductors~\cite{buks98, sprinzak00, avinun04, neder07}.
In these experiments, a quantum point contact (QPC) was used as a
WP detector by probing the charge of a single electron at a nearby
charge qubit composed of a quantum dot~\cite{buks98,sprinzak00,avinun04}
or of a ballistic two-path conductor~\cite{neder07}.
The particular setup we consider here is schematically drawn in Fig.~1:
a charge qubit interacting with a QPC beam collider having four (two input
and two output) electrodes.
It has been well understood that the dephasing of the qubit is caused by
the charge detection when uncorrelated electrons are
injected from one of the
source electrodes and partitioned by the
QPC~\cite{aleiner97, gurvitz97, levinson97,
hacken98, stodolsky99, buttiker00, averin05, kang05, khym06}.

In this Letter, we report our investigations of the dephasing properties of
the qubit when the detector electrons, injected from the two
input electrodes, collide
at the QPC. Interestingly, we find that the
dephasing is suppressed (i.e., the interference is preserved) as a
result of the two-particle collision. When the two electrons,
injected from the two different inputs, collide at the QPC, Fermi statistics
leads to antibunching of electrons. As a result, two electrons coming from
the two input leads are
transferred to the two different output leads because of the Pauli's
exclusion principle manifested in two-particle interference.
The antibunching of
electrons makes it impossible, even in principle, to extract the information
of the charge state despite the charge sensitivity of the scattering
coefficients of the QPC detector. We argue that this shows the nonlocal
nature of dephasing.
We also discuss the case of entangled electron pairs injected from the
two input electrodes. Our observations indicate that the information
itself, rather than
disturbance, indeed brings about
a particle-like behavior of the qubit.

The system under consideration is composed of a charge qubit interacting
with a QPC detector having four electrodes (Fig.~1(a)).
This kind of detector can
be constructed with the quantum Hall bar and split gates as schematically
drawn in Fig.~1(b). We could also utilize the interference of the two
output beams (dashed lines of Fig.1(a,b)) for a phase-sensitive
charge detection.
Constructing interference~\cite{ji03} far away
from the qubit does not influence dephasing of the qubit,
but controls the efficiency of detection~\cite{averin05,kang07}.
The electron spin is neglected at this point of discussion.
(Charge detection with spin-entangled electrons is discussed
later.)
The qubit, composed of two states, namely $|0\rangle$ and
$|1\rangle$,  may either be a quantum
dot~\cite{buks98,sprinzak00} or be a two-path interferometer~\cite{neder07}.
Creation (Annihilation) of an electron at each electrode $x$
($\in\alpha,\beta,\gamma, \delta$) is represented by the operator
$c_x^\dagger$ ($c_x$).
The characteristics of the scattering of an electron at the
QPC is accounted for
by the scattering matrix
\begin{equation}
 S_i = \left( \begin{array}{ll}
                r_i & t_i' \\
                t_i & r_i'
           \end{array}  \right) \;,
 \label{eq:S-mtx}
\end{equation}
depending on the charge state $i$ ($\in0,1$) of the qubit,
which transforms the electron operators as
\begin{equation}
 \left( \begin{array}{c} c_\gamma \\ c_\delta
        \end{array} \right) = S_i
 \left( \begin{array}{c} c_\alpha \\ c_\beta
        \end{array} \right) .
\end{equation}

\begin{figure}[b]
\includegraphics[width=7.0cm]{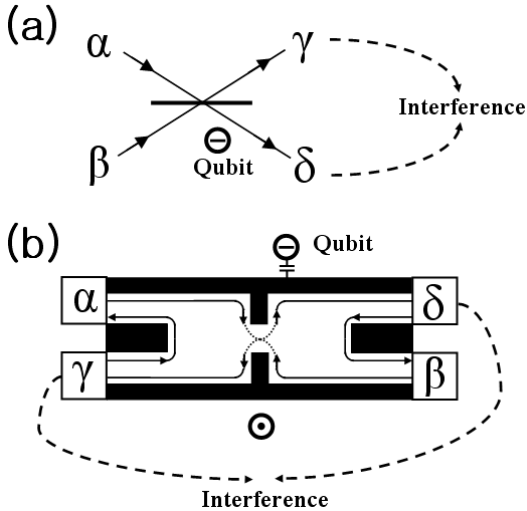}
\caption{\label{BS} (a) A schematic diagram of a charge qubit electrostatically
coupled to a detector with a beam splitter and four (two input and two output)
electrodes, (b) which can be realized by using the quantum Hall bar and
quantum point contact.}
\end{figure}

Charge detection and dephasing induced by the detection have been extensively
studied previously when one of the input electrodes injects uncorrelated
electrons~\cite{aleiner97,gurvitz97,levinson97,hacken98,stodolsky99,buttiker00,averin05,kang05,khym06}.
In our setup of Fig.~1, this situation
can be reproduced when one of the input electrodes is biased and all the other
electrodes are grounded. First we briefly review the detector-induced
dephasing in this case. When an electron is injected from input $\alpha$,
the wave function, $|\psi\rangle$, is composed of the
individual wave functions of the qubit, $a_0|0\rangle+a_1|1\rangle$,
and the detector state, $c_\alpha^\dagger|F\rangle$.
($|F\rangle$ denotes Fermi sea of all the electrodes with
energy lower than zero.) It evolves as
\begin{equation}
 \left( a_0|0\rangle+a_1|1\rangle \right) \otimes
  c_\alpha^\dagger|F\rangle \rightarrow
 a_0|0\rangle\otimes|\chi_0\rangle + a_1|1\rangle\otimes|\chi_1\rangle ,
\label{eq:psi}
\end{equation}
where $|\chi_i\rangle=(r_i c_\gamma^\dagger +
t_i c_\delta^\dagger)|F\rangle$ ($i=0,1$).
This results in an evolution of the reduced density matrix
$\rho$ of the qubit,
$\rho = Tr_{det}|\psi\rangle\langle\psi|$, obtained by tracing over
the detector states of Eq.(\ref{eq:psi}):
\begin{equation}
 \rho_{ij} = a_ia_j^* \rightarrow a_ia_j^*\langle\chi_j|\chi_i\rangle
  = a_ia_j^* (r_ir_j^*+t_it_j^*).
\label{eq:rho1}
\end{equation}
This leads to suppression of $\rho_{ij}$ for $i\ne j$,
which gives rise to dephasing of the qubit state uppon continuous
injection of detector electrons.

Now, let us consider the situation when electrons are injected from both
of the input electrodes $\alpha$ and $\beta$ so that two electrons collide
at the QPC.
In this case, the initial detector state, $c_{\alpha}^\dag
c_{\beta}^\dag|F\rangle$, evolves into
\begin{displaymath}
|\chi_i\rangle = (r_i c_{\gamma}^\dag + t_i c_{\delta}^\dag)(t'_i
c_{\gamma}^\dag + r'_i c_{\delta}^\dag)|F\rangle,
\end{displaymath}
where $i$ denotes the charge state of the qubit (being
$i=0$ or $i=1$). Considering Fermi statistics,
$\{c_{x},
c_{y}^\dag\}=\delta_{xy}$, we find
\begin{equation}
|\chi_i\rangle= (r_i r'_i - t_i t'_i)
    c_{\gamma}^\dag c_{\delta}^\dag |0\rangle
  = e^{i\theta_i} c_{\gamma}^\dag c_{\delta}^\dag |F\rangle,
\label{eq:chi2}
\end{equation}
where $\theta_i=\arg{(r_ir_i')}=\arg{(t_it_i')}+\pi$ is the global phase
of $S_i$. The latter
equality of Eq.~(\ref{eq:chi2}) is a result of the unitarity
of $S_i$.
As a result of two-particle interference and Fermi statistics,
the detector state of Eq.(\ref{eq:chi2}) has only one
particular possibility that each electron propagates into different
output lead.  This implies that the evolution of the density matrix
of the qubit is given as
\begin{equation}
 \rho_{ij} = a_ia_j^* \rightarrow a_ia_j^*\langle\chi_j|\chi_i\rangle
  = a_ia_j^* e^{i(\theta_i-\theta_j)} .
\label{eq:rho2}
\end{equation}
{\em Apparently, the two-particle collision in the detector
does not reduce the interference, unlike the case of single-particle
scattering}.

To be specific, we consider a general case with many
electrons injected from the two input electrodes
$\alpha$ and $\beta$ biased by
$V_\alpha$ and $V_\beta$ ($V_\alpha\geq V_\beta>0$), respectively.
The two output electrodes are grounded ($V_\gamma=V_\delta=0$).
The state of composite qubit-detector system initially given as
\begin{equation}
 \left( a_0|0\rangle+a_1|1\rangle \right) \otimes
\left[  \prod_{0<\varepsilon\leq eV_\beta} c_\alpha^\dagger(\varepsilon)
                            c_\beta^\dagger(\varepsilon)
  \prod_{eV_\beta<\varepsilon'\leq eV_\alpha} c_\alpha^\dagger(\varepsilon)
   |F\rangle \right] ,
 \end{equation}
gets entangled upon interaction between the two subsystems.
Successive events of scattering give the time dependence of the reduced
density matrix as
\begin{equation}
 \log{[\rho_{ij}(t)]} = \log{[\rho_{ij}(0)]} + \sum_\varepsilon
         \log{\lambda_{ij}(\varepsilon)},
\label{eq:rhoN}
\end{equation}
where $\lambda_{ij}(\varepsilon)$ corresponds to the indisitinguishability
parameter
of the detector electrons with energy $\varepsilon$ (just
as in $\langle\chi_j|\chi_i\rangle$ for the simpler cases in
Eqs.~(\ref{eq:rho1},\ref{eq:rho2})). We find
\begin{equation}
  \lambda_{ij}(\varepsilon) = \left\{
    \begin{array}{ll}
     e^{i(\theta_i-\theta_j)} & 0 <\varepsilon < eV_\beta \\
     r_j^*r_i+ t_j^*t_i & eV_\beta <\varepsilon < eV_\alpha \\
    1 & \mbox{\rm otherwise}
    \end{array} \right.   \,.
\end{equation}
At time $t\gg h/eV_\alpha$ where the energy-time phase space is much larger
than $h$, the summation $\sum_\varepsilon$ can be
replaced by $t\int d\varepsilon/h$. In this limit, we obtain
$|\rho_{01}(t)| = |\rho_{01}(0)|\exp{(-\Gamma_d t)}$ with
the dephasing rate $\Gamma_d$ given by
\begin{equation}
 \Gamma_d = -\int \frac{d\varepsilon}{h}
   \log{\left|\lambda_{01}(\varepsilon)\right|},
\end{equation}
and we get
\begin{equation}
  \Gamma_d = -\frac{e|V_\alpha-V_\beta|}{h} \log{|r_0r_1^*+t_0t_1^*|} .
\label{eq:deph}
\end{equation}
In the weak coupling limit ($r_0r_1^*+t_0t_1^*\sim1$), the dephasing
rate can be expanded in terms of the
change in the transmission probability,
$\Delta T = |t_{1}|^{2} - |t_{0}|^{2}$,
and the change in the relative scattering phase
$\Delta\phi=arg(t_{1}/r_{1})-arg(t_{0}/r_{0})$. This expansion results in
\begin{subequations}
\begin{eqnarray}
 \Gamma_{d} &=& \Gamma_{T} + \Gamma_{\phi},  \\
 \Gamma_{T} &=& \frac{e|V_\alpha-V_\beta|}{h}\frac{(\Delta  T)^{2}}{8T(1-T)},
  \\
 \Gamma_{\phi} &=& \frac{e|V_\alpha-V_\beta|}{2h} T(1-T)(\Delta\phi)^{2},
\end{eqnarray}
\label{eq:deph-weak}
\end{subequations}
where $T=( |t_1|^2 + |t_0|^2 )/2$.

Eqs.(\ref{eq:deph},\ref{eq:deph-weak}) are our central result.
When only one of the input electrodes $\alpha$ injects electrons,
that is for $V_\alpha>0$ and $V_\beta=0$,
Eqs.(\ref{eq:deph},\ref{eq:deph-weak}) correspond to the previously studied
dephasing rate through partitioning the uncorrelated
electrons~\cite{aleiner97, gurvitz97, levinson97, hacken98, stodolsky99,
buttiker00, averin05, kang05, khym06}.  Turning
on the bias of the other input $\beta$ results in the decrease of the dephasing
rate in spite of the increasing number of detector electrons.
For identical biases,
$V_\alpha=V_\beta$, the dephasing rate vanishes. This intriguing result
originates from the two-electron collisions which do not reduce the
interference in the qubit, and can be understood as follows.
As shown in Eq.(\ref{eq:chi2}), two electrons cannot scatter into
the same output lead because of Fermi statistics. This ``antibunching"
makes the transport noiseless~\cite{liu98}.
Therefore, output currents at lead $\gamma$
and $\delta$ are insensitive to the charge state of the qubit ($\Delta T$
in the scattering coefficients plays no role). Furthermore,
the phase sensitivity $\Delta\phi$ does not either affect the detector
in any noticeable way when an interferometer is constructed between
the two output leads. Therefore,
{\em charge detection is impossible, even in principle,
through the two-electron collision.}

Our result indicates the {\em nonlocality of dephasing}.
The origin
of dephasing can be interpreted either by information acquisition
in the detector, or by back action of the detector causing the random
fluctuation of the phase in the qubit~\cite{stern90}. The
`back-action dephasing' is often identified with ``momentum
kick" or local ``disturbance" imposed by the uncertainty
principle~\cite{feynman65}.
In the ``back-action" interpretation, one might be tempted to assume
a picture that the
local Coulomb interaction exerts force (or a momentum kick) to the qubit
leading to uncertainty of the phase. However, our result
shows that this naive picture should be discarded.
Injecting additional electrons at lead $\beta$ does not
affect the scattering matrix of Eq.~(\ref{eq:S-mtx}) as long as
the lead $\beta$
is far apart from the qubit. If the local disturbance were the only origin
of dephasing, increasing $V_\beta$ would always monotonically
raise the dephasing rate due to the increment of detector electrons. However, as
we find above, the two-electron collision does not contribute to dephasing
in spite of charge sensitivity of the scattering matrix,
and it verifies the nonlocality of dephasing. We emphasize
that the particle-like behavior of the qubit emerges only when the
charge state information could be acquired in the detector, even if
it could be done only in principle~\cite{khym06,kang07}.



Next, let us consider injection of spin-entangled
electrons from the two input leads identically biased with $V$ (Fig.~2).
Some possible implementations of the spin-entangled electrons in
solid-state circuits are found in Ref.~\onlinecite{burkard07}.
The ``entangler" injects spin-entangled electrons to the leads $\alpha$ and
$\beta$.
The scattering matrix at the QPC is assumed to be spin-independent and is
given by Eq.~(\ref{eq:S-mtx}).
The injected entangled triplet(singlet), prior to scattering at the
QPC, is written as~\cite{burkard00}
\begin{equation}
 \frac{1}{\sqrt{2}}
(c^{\dag}_{\alpha\uparrow} c^{\dag}_{\beta\downarrow} \pm
c^{\dag}_{\alpha\downarrow} c^{\dag}_{\beta\uparrow} ) |F\rangle,
\label{eq:ent}
\end{equation}
where
$\uparrow$ and $\downarrow$ represent the spin state of an electron.
The $+(-)$ sign in Eq~(\ref{eq:ent}) corresponds to the triplet(singlet)
state.
Upon collision at the QPC it is reduced to
the qubit-charge-dependent detector state
$|\chi_i^{t(s)}\rangle$ given by
\begin{eqnarray*}
|\chi_i^{t(s)}\rangle = \frac{1}{\sqrt{2}}    \{ ( r_i
c^{\dag}_{\gamma\uparrow}  + t_i c^{\dag}_{\delta\uparrow} ) (
t_{i}' c^{\dag}_{\gamma\downarrow}  + r_{i}'
c^{\dag}_{\delta\downarrow} ) \pm  \nonumber \\
  ( r_i c^{\dag}_{\gamma\downarrow} + t_i c^{\dag}_{\delta\downarrow} ) (
t_{i}' c^{\dag}_{\gamma\uparrow} + r_{i}' c^{\dag}_{\delta\uparrow}
) \} ~|F\rangle.
\end{eqnarray*}
Again, Fermi statistics,
$\{c_{i\sigma}, c_{j\sigma'}^\dag\}=\delta_{ij}
\delta_{\sigma\sigma'}$,
is crucial in characterizing the detector properties.
We find that the triplet state is simplified as
\begin{equation}
|\chi_i^t\rangle = \frac{1}{\sqrt{2}} e^{i\theta_i}
(c^{\dag}_{\gamma\uparrow} c^{\dag}_{\delta\downarrow} +
c^{\dag}_{\gamma\downarrow} c^{\dag}_{\delta\uparrow} ) |F\rangle,
\label{eq:chit}
\end{equation}
which leads to the indistinguishability parameter $\lambda_{ij}$ of
Eq.~(\ref{eq:rhoN}) as
\begin{equation}
  \lambda_{ij}(\varepsilon) = \left\{
    \begin{array}{ll}
     e^{i(\theta_i-\theta_j)} & 0 <\varepsilon < eV \\
     1 & \mbox{\rm otherwise}
    \end{array} \right.   \,.
\label{eq:lambda_t}
\end{equation}
As we find from Eqs.~(\ref{eq:rhoN},\ref{eq:lambda_t}),
the dephasing rate vanishes when the input electrodes inject
triplet pairs just as in the collision of independent electrons.
This is again due to the antibunching of the orbital wave function of
electrons which provides
noiseless beam upon collision. The orbital wave function of the
triplet state is
antisymmetric under exchange, and its statistics is equivalent to
that of the independent fermions~\cite{burkard00}.

\begin{figure}[b]
\includegraphics[width=7.0cm]{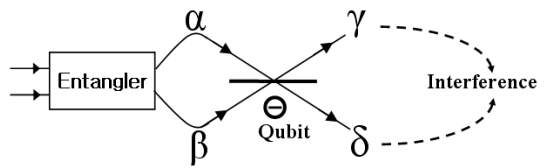}
\caption{\label{EntangledE} A schematic diagram of a charge qubit
and a detector that injects
spin-entangled electrons.
 }
\end{figure}

In contrast, the orbital wave function of the singlet is symmetric under
two-particle exchange. Therefore we expect the detection property to
be equivalent to that of bosons.
Indeed, collision of the singlet pair at the QPC leads to
the detector state of the form
\begin{eqnarray}
|\chi_i^s\rangle &=& \sqrt{2} \left[
  r_i t_{i}'c^{\dag}_{\gamma\uparrow} c^{\dag}_{\gamma\downarrow}
  + t_i r_{i}'c^{\dag}_{\delta\uparrow} c^{\dag}_{\delta\downarrow} \right.
    \nonumber \\
  & & \left. +\frac{1}{2}(t_it_i'+r_ir_i')
    ( c^{\dag}_{\gamma\uparrow} c^{\dag}_{\delta\downarrow}
    + c^{\dag}_{\delta\uparrow} c^{\dag}_{\gamma\downarrow}) \right] |F\rangle.
  \label{eq:chis}
\end{eqnarray}
This singlet detector state, unlike those of the triplet
(Eq.~(\ref{eq:chit})) and of the two independent electrons
(Eq.~(\ref{eq:chi2})), has a ``bunching" property,
which enhances the
shot noise~\cite{burkard00}. The bunching is perfect for the symmetric
partitioning at the QPC (that is $|t_i|=|r_i|=1/\sqrt{2}$)
where $t_it_i'+r_ir_i'= e^{i\theta_i}(|r_i|^2-|t_i|^2)=0$. In this
case, the two electrons
are always found at the same lead ($\gamma$ or $\delta$). Moreover,
this bunching enhances the charge sensitivity of the detector.
For the detector injecting singlet pairs,
we find that the dephasing rate $\Gamma_d^s$ is given as
\begin{equation}
  \Gamma_d^s = -\frac{2eV}{h} \log{|\lambda_{01}^s|} ,
\label{eq:Gammas}
\end{equation}
where $\lambda_{01}^s = \langle\chi_1^s|\chi_0^s\rangle =
2(r_1^* t_1'^* r_0t_0'+t_1^* r_1'^* t_0r_0')
 + (t_1^* t_1'^*+r_1^* r_1'^*)(t_0t_0'+r_0r_0')$ is the indistinguishability
factor for a singlet pair. Factor $2$ in the right hand side of
Eq.~(\ref{eq:Gammas})
comes from the spin degeneracy, which was not taken into account in
Eq.~(\ref{eq:deph}).
In the weak measurement limit, $\Gamma_d^s$ is given by an algebraic sum of
the two different contribution: $\Gamma_d^s = \Gamma_T^s+\Gamma_\phi^s$,
where the current-sensitive ($\Gamma_T^s$)
and the phase-sensitive ($\Gamma_\phi^s$) contributions are given as
\begin{subequations}
\begin{eqnarray}
 \Gamma_T^s &=& \frac{eV}{h}\frac{(\Delta  T)^{2}}{T(1-T)},
  \\
 \Gamma_\phi^s &=& \frac{eV}{h} 4T(1-T)(\Delta\phi)^{2},
\end{eqnarray}
\end{subequations}
The dephasing rate is now enhanced (by eight times) compared to the
case with only one
electrode injecting uncorrelated electrons ($V_\beta=0,V_\alpha=V$ in
Eq.~(\ref{eq:deph-weak})).  Taking into
account the simultaneous injection from the two inputs and the spin
degeneracy, the number of injected electrons for a given time is
four times larger in the case of injecting singlet states.
This means that the charge
sensitivity of the singlet pairs is twice as compared
to that of the uncorrelated single electrons. This originates from the
bunching behavior of the orbital
wave function. It is noteworthy that this scheme may be utilized to
achieve more precise charge detection~\cite{yurke86}.


In conclusion, we have analyzed the properties of charge detection in
a QPC when the electrons from different inputs collide.
We have found that the properties of dephasing are determined by the
statistics of the incident electrons, and demonstrated
the nonlocality of dephasing. This verifies that,
while the dephasing is directly related to the which-path information
in general, it cannot be simply understood in terms of local disturbance
that washes out the coherence.

\acknowledgements%
This work was supported by the Korea Research Foundation
(KRF-2005-070-C00055, KRF-2006-331-C00116).


\end{document}